\documentclass[12pt]{article}
\usepackage{amssymb,amsfonts}
\usepackage{amsmath}
\usepackage{epsf,epsfig}
\pdfoutput=1
\begin{document}
\baselineskip 18pt

\title{The competition between low-temperature kinks and magnons at the vicinity of the deconfinement transition point in 1D easy-axis XXZ ferromagnet}
\author{P.~N.~Bibikov}

\maketitle

\vskip5mm

\begin{abstract}
Studying the ordered phases and quantum supercritical low-temperature regime at the vicinity of the deconfinement transition point in 1D easy-axis XXZ ferromagnet, we suggest their interpretations according to the corresponding dominant lowest-energy excitations. We show, that the two ordered phases are governed by magnons, while the quantum supercritical regime is governed by kinks. Within this framework the Ising model is treated in detail.
\end{abstract}

\maketitle
\vskip20mm

\section{Introduction}

The unconventional, different from the Landau-Ginsburg-Wilson-Fisher, paradigm of phase transition was suggested in \cite{1}. It describes critical points, separating {\it two} phases characterized by conventional "confining" order parameters. The progress in this direction within the last 20 years is presented in the two recent reviews \cite{2,3}. A typical example of such transition is the alternation of magnetization, when as
it was noted in \cite{3}, "magnons with integer spins dominate the excitations in the ordered phases on both sides of the transition, whereas
at the critical point, the excitations turn out to be fractionalized as free spinons with half-integer spins". Within this interpretation, magnons
in a ferromagnetic chain are treated as confined two-spinon pairs. This picture is very similar to the confinement of kinks (spinons) into mesons
(magnons) in a 1D XXZ antiferromagnet under the action of a staggered magnetic field \cite{4}. Another example is the
transition between the rung-dimerized and the rung-polarized phases in a spin ladder \cite{5}.

In the present paper we consider, the mentioned above, alternation magnetization transition in an easy-axis XXZ ferromagnetic chain,
corresponding to the Hamiltonian
\begin{equation}\label{ham}
\hat H^{\rm XXZ}=-\sum_{n=-M}^M\Big[J_z\Big({\bf S}^z_n{\bf S}^z_{n+1}-\frac{1}{4}\Big)
+\frac{J_{xy}}{2}\Big({\bf S}^+_n{\bf S}^-_{n+1}+{\bf S}^-_n{\bf S}^+_{n+1}\Big)
+h{\bf S}^z_n\Big],
\end{equation}
where ${\bf S}^j$ ($j=+,-,z$) are the usual spin-1/2 operators. We postulate ${\bf S}_{\pm(M+1)}={\bf S}_{\mp M}$ for {\it periodic} and
${\bf S}_{\pm(M+1)}=0$ for {\it open} chains. Also it will be assumed that
\begin{equation}\label{Jz>0}
J_z>0,
\end{equation}
and
\begin{equation}\label{|Delta|}
|\Delta|>1,
\end{equation}
where the, so called, anisotropy parameter is
\begin{equation}\label{Delta}
\Delta\equiv\frac{J_z}{J_{xy}}.
\end{equation}

At $h\geq0$ the corresponding to \eqref{ham} ground state
\begin{equation}\label{vac+}
|\emptyset_+\rangle=\prod_{n=-M}^M|\uparrow\rangle_n,
\end{equation}
is ferromagnetically polarized. The same is true for $h\leq0$, when the ground state
\begin{equation}\label{vac-}
|\emptyset_-\rangle=\prod_{n=-M}^M|\downarrow\rangle_n,
\end{equation}
has the opposite polarization. At $h=0$ both \eqref{vac+} and \eqref{vac-} are the ground states of \eqref{ham}.

At $T=0$ the alternation of $h$ results in distinct phase transition between \eqref{vac+} and \eqref{vac-} under which the system passes through the
intermediate point $h=0$.
The latter, at $T>0$, expands into a finite area, associated with the quantum supercritical regime (QSR) and
corresponding to a continuous crossover with gradual change of magnetization \cite{6,7}. The typical
$h-T$ phase diagram is presented in the Fig. 1b of \cite{6} and includes three segments, the order I (grounded on \eqref{vac-}), the order II
(grounded on \eqref{vac+}) and, between them, the (intermediate between \eqref{vac-} and \eqref{vac+}) QSR phase, where the system behaves very similar to a super critical liquid.

Since, in all the mentioned regimes, the system is gapped, it seems natural to suppose, that their low-temperature thermodynamical behaviors at
\begin{equation}\label{lowT}
T<T_{\rm gap}\equiv\frac{E_{\rm gap}}{k_{\rm B}},
\end{equation}
($E_{\rm gap}$ is the gap energy) are governed by the corresponding lowest energy spectrums. Practically, this means that almost all the low-temperature thermodynamical parameters may be obtained, with satisfactory accuracy, by manipulations only with the ground states and the lowest energy excitations.  At $|h|>0$ such excitations are magnons, while at $h=0$ they are kinks (spinons). Nevertheless, as it will be shown in the present paper,
at $T>0$ kinks should govern the low-temperature thermodynamics not only directly on the line $h=0$ but almost within the whole QSR. Our phenomenological argumentation on this assumption is the following.

At $|h|>0$ the lowest-energy excitations, for the corresponding to \eqref{ham}, {\it periodic} chain are magnons \cite{8}
\begin{equation}\label{magn}
|{\rm magn};k\rangle_{\pm}=\sum_{n=-M}^M{\rm e}^{ikn}{\bf S}^{\mp}_n|\emptyset_{\pm}\rangle,\qquad k\in[0,2\pi],\qquad h=\pm|h|,
\end{equation}
with energies
\begin{equation}\label{Emagn}
E_{\rm magn}(k)=|h|+J_z-J_{xy}\cos{k}.
\end{equation}
At $h=0$ the lowest-energy {\it open} chain excitations are kinks and antikinks with equal energies \cite{9}
\begin{equation}\label{Ekink}
E_{\rm kink}=\frac{J_z}{2}\sqrt{1-\frac{1}{\Delta^2}}.
\end{equation}
Exact representations for the kink and antikink states are rather transparent only for the Ising model, corresponding to $J_{xy}=0$. In this case
\begin{eqnarray}
&&|{\rm kink};n\rangle^{\rm Is}=\prod_{m=-M}^n\otimes|\uparrow_m\rangle\otimes\prod_{m=n+1}^M\otimes|\downarrow_m\rangle,\nonumber\\
&&|{\rm antikink};n\rangle^{\rm Is}=\prod_{m=-M}^n\otimes|\downarrow_m\rangle\otimes\prod_{m=n+1}^M\otimes|\uparrow_m\rangle,
\end{eqnarray}
where $|\uparrow_m\rangle$ and $|\downarrow_m\rangle$ are up and down spin states in the space ${\mathbb C}^2$, attached to the $m$-th site.

At $|h|>0$ and $M\rightarrow\infty$ the energy of both kink and antikink, with respect to the ground states either \eqref{vac+} or \eqref{vac-}, becomes infinite, however the energy of a kink-antikink pair
\begin{equation}
|{\rm kink-antikink};n_1,n_2\rangle^{\rm Is}=\prod_{m=-M}^{n_1}\otimes|\uparrow_m\rangle\otimes
\prod_{n_1+1}^{n_2}|\downarrow_m\rangle\otimes\prod_{m=n_2+1}^M\otimes|\uparrow_m\rangle,
\end{equation}
with respect to \eqref{vac+}, remains finite at $|h|>0$. The same is true for the antikink-kink pair
\begin{equation}
|{\rm antikink-kink};n_1,n_2\rangle^{\rm Is}=\prod_{m=-M}^{n_1}\otimes|\downarrow_m\rangle\otimes
\prod_{n_1+1}^{n_2}|\uparrow_m\rangle\otimes\prod_{m=n_2+1}^M\otimes|\downarrow_m\rangle,
\end{equation}
energy with respect to \eqref{vac-}. In both the cases, these energies
\begin{equation}\label{Epair}
E_{\rm kink-antikink}(n_1,n_2)=E_{\rm antikink-kink}=J_z+|h|{\tt L},
\end{equation}
depend linearly on the length parameter
\begin{equation}
{\tt L}=n_2-n_1.
\end{equation}

At $h=0$ and $T<T_{\rm gap}$ the system behaves as a thermally excited rare gas of kinks and antikinks in the deconfined phase. The corresponding Hilbert space ${\cal H}$ decomposes on the direct sum of topological sectors
\begin{equation}\label{decomp}
{\cal H}={\cal H}_{\uparrow\uparrow}\oplus{\cal H}_{\uparrow\downarrow}\oplus{\cal H}_{\downarrow\uparrow}\oplus{\cal H}_{\downarrow\downarrow},
\end{equation}
where for $|\psi_{c^{(L)}c^{(R)}}\rangle\in{\cal H}_{c^{(L)}c^{(R)}}$ one has for some $-M\leq M_L<M_R\leq M$
\begin{equation}
|\psi_{c_Lc_R}\rangle=\prod_{m=-M}^{M_L}\otimes|c_m^{(L)}\rangle\otimes\dots\otimes\prod_{m=M_R}^M\otimes|c_m^{(R)}\rangle.
\end{equation}
Under even a subtle increase of $|h|$ the decomposition \eqref{decomp} essentially reduces. Namely, at $h>0$ its right side turns into
${\cal H}_{\uparrow\uparrow}$ and should be treated as a rare gas of kink-antikink pairs. Contrary, at $h<0$ it turns into the, corresponding to ${\cal H}_{\downarrow\downarrow}$, rare antikink-kink pair gas. In both the cases, the system remains in the QSR.

According to \eqref{Epair}, a further increase of $|h|$ results in the decrease of the pair lengths up to their minimal values ${\tt L}=1$.
Since, such a pair is exactly a magnon, the system, on this stage, turns into one of the ordered confined phases.

However, if on this stage one to begin to heat the chain, then the pair lengths, as well as the pair density, will simultaneously increase. Under this process, the system will return into the QSR, however now as a dense gas of antikink-kink (or kink-antikink if $h<0$) pairs.

Following the presented above physical picture, we suppose, that the $T<T_{\rm gap}$ mathematical description of the ordered phases should be done on the basis of the magnon spectrum, while for the QSR one should use kinks.

\section{The kink and magnon fingerprints in the low-temperature asymptotic of the free energy density for XXZ chain}

The cluster, or fugacity expansion (the term "sector expansion" seems more convenient for the author in the context of spin chains)
is an effective, direct (not based on integrability) theoretical approach for evaluation of low-temperature thermodynamical properties \cite{10,11}
under a very clear physical interpretation.
Within it, the low-temperature asymptotic for the free energy density of the system
\begin{equation}\label{f}
f(T,h)\equiv-\frac{1}{\beta}\lim_{N\rightarrow\infty}\frac{1}{N}\log{\rm Tr}({\rm e}^{-\beta\hat H}),\qquad N\equiv 2M+1,
\end{equation}
is transformed into the infinite sum
\begin{equation}\label{fsect}
f(T,h)=\varepsilon_0(h)+\sum_{j=1}^{\infty}f_j(T,h),
\end{equation}
where $\varepsilon_0(h)$ is the ground state energy density, while each $f_j(T,h)$ may be obtained by utilization of only the $l$-particle states energies with $l\leq j$ \cite{12}. For example,
\begin{equation}\label{fclust}
f_1(T,h)=-k_BT\lim_{N\rightarrow\infty}\frac{1}{N}\sum_{\mu}{\rm e}^{-\beta E_{\mu}}+o({\rm e}^{-\beta E_{\rm gap}}),\qquad
\beta=\frac{1}{k_{\rm B}T},
\end{equation}
where the parameter $\mu$ enumerates the lowest-energy excitation branch. Since the latter depends on the phase of the system, its
correct identification is the main state of art.
However, for rather simple models
this procedure usually is readily solvable. Namely, the assumption that the lowest-energy excitations in the ordered phase are magnons, automatically
yields \cite{10,11}
\begin{equation}\label{ford}
\varepsilon^{\rm magn}_0(h)=-\frac{|h|}{2},\qquad f^{\rm magn}_1(T,h)=-\frac{k_BT}{2\pi}\int_0^{2\pi}dk{\rm e}^{-\beta E_{\rm magn}(k)}.
\end{equation}

At the same time, in the QSR \eqref{fclust} there should be
\begin{equation}\label{fQSR}
\varepsilon_0^{\rm QSR}(h)=0,\qquad f^{\rm QSR}_1(T)=-k_BT{\rm e}^{-\beta E_{\rm kink}}.
\end{equation}

Though the derivations of \eqref{ford} and \eqref{fQSR} from \eqref{fclust} are rather elementary, it is not clear, what formula should
be utilized at concrete values of $T$ and $h$. This, however, my be done under comparing them with results, obtained by alternative methods,
theoretical or numerical. The only known for the author result in this direction is the low-temperature formula
\begin{equation}\label{JB}
f(T,h)\approx\sigma h-\sqrt{\frac{h^2}{4}+(k_BT)^2{\rm e}^{-\beta|J_{xy}|\sqrt{\Delta^2-1}}}
-\frac{k_BT}{2\pi}\int_0^{2\pi}dk{\rm e}^{-\beta E_{\rm magn}(k)},
\end{equation}
presented in Eq. (2.21) of \cite{13} (under the assumption $|J_{xy}|=1$) on the within the thermodynamical Bethe ansatz approach \cite{13}
(in \cite{13} $\sigma$ is the magnetization per spin).
Since \eqref{JB} was derived in \cite{13} only within physical order of strictness, its rigorous proof remains an actual problem.

Suggesting, however, that the formula \eqref{JB} is correct, one may be proved by direct comparison between their terms
that under the condition \cite{13}
\begin{equation}\label{Delta5/3}
\Delta>\frac{5}{3},
\end{equation}
thermodynamical behaviors of the system in the two following scaling regimes
\begin{equation}\label{gg}
\beta|h|\gg1,
\end{equation}
and
\begin{equation}\label{ll}
\beta|h|\ll1,
\end{equation}
at the vicinity of the supercritical point $T=0$, $h=0$ should be quite different.

In the regime \eqref{gg}, when
\begin{equation}
\frac{h^2}{4}\gg(k_BT)^2{\rm e}^{-\beta|J_{xy}|\sqrt{\Delta^2-1}},
\end{equation}
\eqref{JB} reduces to
\begin{equation}\label{JBm}
f(T,h)\approx\sigma h
-\frac{k_BT}{2\pi}\int_0^{2\pi}dk{\rm e}^{-\beta E_{\rm magn}(k)},
\end{equation}
so that the low-temperature thermodynamics is governed by magnons.

In the opposite regime \eqref{ll}, when
\begin{equation}
\frac{h^2}{4}\ll(k_BT)^2{\rm e}^{-\beta|J_{xy}|\sqrt{\Delta^2-1}},
\end{equation}
\eqref{JB} reduces to
\begin{equation}\label{JBk}
f(T,h)\approx-k_BT{\rm e}^{-\beta|J_{xy}|\sqrt{\Delta^2-1}/2}.
\end{equation}
The latter, according to \eqref{Ekink} and \eqref{Delta}, shows that \eqref{ll} is governed by kinks.

Summarizing, we suggest that \eqref{JB} gives the strong evidence, that the scaling region \eqref{gg} should be governed by magnons
(and hence corresponds to the ordered phase). At the same time, the phase \eqref{ll} should be governed by kinks (and, hence, correspond to QSR).

\section{The low-temperature behavior of the ferromagnetic Ising chain}

\subsection{The free energy density}

At $J_{xy}=0$, \eqref{ham} turns into the Ising Hamiltonian $\hat H^{\rm Is}$, with the free energy density \cite{14}
\begin{equation}\label{fIs0}
f_{\rm Is}(T,h)=-k_BT\log{\Big[\cosh{\frac{\beta h}{2}}+\sqrt{\sinh^2{\frac{\beta h}{2}}+{\rm e}^{-\beta J_z}}\Big]}.
\end{equation}
The low-temperature expression for \eqref{fIs0} in the region \eqref{gg} (denoted here as $f_{\rm Is}^{\rm magn}(T,h)$) may be obtained by the use of the cluster (sector) expansion
\cite{10,11}
\begin{equation}\label{fIsmagn}
f_{\rm Is}^{\rm magn}(T,h)\approx\varepsilon_{\rm Is\,0}^{\rm magn}(h)+f_{\rm Is\,1}^{\rm magn}(T,h)+f_{\rm Is\,2}^{\rm magn}(T,h),
\end{equation}
where
\begin{equation}\label{f0}
\varepsilon_{\rm Is\,0}^{\rm magn}(h)=-\frac{|h|}{2},
\end{equation}
corresponds to the ground state energy density, while
\begin{equation}\label{f1}
f_{\rm Is\,1}^{\rm magn}(T,h)=-k_BT\lim_{N\rightarrow\infty}\frac{Z_{N\,1}^{\rm magn}(T,h)}{N},
\end{equation}
and
\begin{equation}\label{f2}
f_{\rm Is\,2}^{\rm magn}(T,h)=-k_BT\lim_{N\rightarrow\infty}\frac{2Z_{N\,2}^{\rm magn}(T,h)-(Z_{N\,1}^{\rm magn}(T,h))^2}{2N},
\end{equation}
give contributions from the one- and two-magnon sectors. Here,
\begin{equation}\label{Z12}
Z_{N\,1}^{\rm magn}(T,h)=N{\rm e}^{-\beta E_{\rm magn}^{\rm Is}},\qquad Z_{N\,2}^{\rm magn}(T,h)=\frac{N(N-3)}{2}{\rm e}^{-2\beta E_{\rm magn}^{\rm Is}}+N{\rm e}^{-\beta E_{\rm bound}^{\rm Is}},
\end{equation}
while
\begin{equation}
E_{\rm magn}^{\rm Is}=|h|+J_z,\qquad E_{\rm bound}^{\rm Is}=2|h|+J_z,
\end{equation}
are the energies of a one-magnon and a two-magnon bound states (a two-magnon scattering state has the energy $2E_{\rm magn}^{\rm Is}$).

Under the substitution of \eqref{Z12}, the formulas \eqref{f1} and \eqref{f2} yield
\begin{equation}
f_{\rm Is\,1}^{\rm magn}(T,h)=-k_BT{\rm e}^{-\beta E_{\rm magn}^{\rm Is}},\qquad
f_{\rm Is\,2}^{\rm magn}(T,h)=-k_BT\Big({\rm e}^{-\beta E_{\rm bound}^{\rm Is}}
-\frac{3{\rm e}^{-2\beta E_{\rm magn}^{\rm Is}}}{4}\Big),
\end{equation}
resulting in the low-temperature expression
\begin{equation}\label{fIsmagn}
f_{\rm Is}^{\rm magn}(T,h)\approx-\frac{|h|}{2}-k_BT\Big({\rm e}^{-\beta E_{\rm magn}^{\rm Is}}+{\rm e}^{-\beta E_{\rm bound}^{\rm Is}}
-\frac{3{\rm e}^{-2\beta E_{\rm magn}^{\rm Is}}}{4}\Big),\qquad\beta|h|\gg1.
\end{equation}
As it is shown in the Appendix, \eqref{fIsmagn} may be also directly derived from \eqref{fIs0}.

Since, at $h=0$ both \eqref{vac+} and \eqref{vac-} are the (zero enery) ground states the cluster (sector) expansion procedure should be slightly modified.
Namely, in this case
\begin{equation}\label{Zkink}
Z_N(T)=Z_{N\,0}^{\rm kink}(T)+Z_{N\,1}^{\rm kink}(T)+Z_{N\,2}^{\rm kink}(T)+\dots,
\end{equation}
where
\begin{eqnarray}
&&Z_{N\,0}^{\rm kink}(T)=2,\qquad Z_{N\,1}^{\rm kink}(T)=2(N-1){\rm e}^{-\beta E_{\rm kink}},\nonumber\\
&& Z_{N\,2}^{\rm kink}(T)=2C_{N-1}^2{\rm e}^{-2\beta E_{\rm kink}}=(N-1)(N-2){\rm e}^{-2\beta E_{\rm kink}},
\end{eqnarray}
are contributions from the (two) ground states, one-kink (antikink) sector and kink-antikink (antikink-kink) sectors.

Following \eqref{Zkink}
\begin{equation}
\log{Z_N^{\rm kink}(T)}=\log{2}+\log{\Big(1+\frac{Z_{N\,1}^{\rm kink}(T)}{2}+\frac{Z_{N\,2}^{\rm kink}(T)}{2}+\dots\Big)}.
\end{equation}
So, utilizing the formula,
\begin{equation}\label{log(1+x)}
\log(1+x)=x-\frac{x^2}{2}+o(x^2),
\end{equation}
one readily gets
\begin{equation}\label{fIskink}
f_{\rm Is}^{\rm kink}(T)\approx f_{\rm Is\,1}^{\rm kink}(T)+f_{\rm Is\,2}^{\rm kink}(T),
\end{equation}
where
\begin{eqnarray}\label{fIskink12}
&&f_{\rm Is\,1}^{\rm kink}(T)=-k_BT\lim_{N\rightarrow\infty}\frac{Z_{N\,1}^{\rm kink}(T)}{2N}=-k_BT{\rm e}^{-\beta E_{\rm kink}},\nonumber\\
&&f_{\rm Is\,2}^{\rm kink}(T)=-k_BT\lim_{N\rightarrow\infty}\frac{4Z_{N\,2}^{\rm kink}(T)-(Z_{N\,1}^{\rm kink}(T,h))^2}{8N}
=\frac{k_BT{\rm e}^{-2\beta E_{\rm kink}}}{2}.
\end{eqnarray}
Alternatively one may obtain \eqref{fIskink}, \eqref{fIskink12}, applying \eqref{log(1+x)} to the $h=0$ reduced formula \eqref{fIs0}
\begin{equation}\label{fIsh=0}
f_{\rm Is}^{\rm red}(T)\equiv f_{\rm Is}(T,0)=-k_BT\log{\Big(1+{\rm e}^{-\beta E_{\rm kink}}\Big)}.
\end{equation}

The plots of $f_{\rm Is}(T,h)$, $f_{\rm Is}^{\rm magn}(T,h)$, $f_{\rm Is}^{\rm red}(T)$ and $f_{\rm Is}^{\rm kink}(T)$ subject to
$J_z=5$ and $h=1$ are presented on Fig.1.

\subsection{The $\bf Gr\ddot uneisen$ ratio}

As it was pointed in \cite{6,15,16,17,18,19}, the $\rm Gr\ddot uneisen$ ratio
\begin{equation}\label{G}
\Gamma_h(T)\equiv\frac{1}{T}\Big(\frac{\partial T}{\partial h}\Big)_S,
\end{equation}
or equivalently
\begin{equation}\label{G=}
\Gamma_h(T)=-\frac{\frac{\partial^2f}{\partial T\partial h}}{T\frac{\partial^2f}{\partial T^2}}.
\end{equation}
is the key parameter, characterizing the magnetocaloric effect at the vicinity of the critical point.
For the Ising model it is \cite{18}
\begin{equation}\label{G=2}
\Gamma_h^{\rm Is}(T)=\frac{(hc+J_zs)(c+Q)^2}{2J_z^2s^2Q+J_z^2c(s^2+Q^2)+h(hc+2J_zs)(c+Q)^2},
\end{equation}
where
\begin{equation}
c=\cosh{\frac{\beta h}{2}},\qquad s=\sinh{\frac{\beta h}{2}},\qquad Q=\sqrt{s^2+{\rm e}^{-\beta J_z}}.
\end{equation}

Implying that the thermodynamical behavior at \eqref{ll} is governed by kinks, it is natural to remove \eqref{G=2} by a reduced formula, corresponding to the model with identically zero magnetic field. A serious problem within this ($h=0$) approach is the definition of the derivative
$\frac{\partial^2f}{\partial T\partial h}$. This may be done however by the use of the formula \cite{18}
\begin{equation}\label{lim}
\lim_{h\rightarrow0}\frac{1}{h}\cdot\frac{\partial^2f}{\partial T\partial h}=-\frac{d\chi(T)}{dT},
\end{equation}
where the zero field magnetic susceptibility
\begin{equation}\label{chi}
\chi(T)\equiv-\frac{\partial^2f}{\partial h^2}\Big|_{h=0},
\end{equation}
may be alternatively obtained from the $h$-independent representation
\begin{equation}\label{chi=}
\chi(T)=\lim_{N\rightarrow\infty}\frac{1}{N}\Big[\frac{{\rm Tr}\Big({\rm e}^{-\beta\hat H_{h=0}}({\bf\hat S}^z)^2\Big)}{Z_N(T)}-
\frac{{\rm Tr}^2\Big({\rm e}^{-\beta\hat H_{h=0}}{\bf\hat S}^z\Big)}{Z^2_N(T)}\Big].
\end{equation}
Here
\begin{equation}
Z_N(T)\equiv{\rm Tr}\Big({\rm e}^{-\beta\hat H^{\rm Is}_{h=0}}\Big),
\end{equation}
is the partition sum associated with the zero field Ising Hamiltonian $\hat H^{\rm Is}_{h=0}$, given by \eqref{ham} under the assumptions $J_{xy}=0$ and $h=0$. Following \eqref{lim}, we suggest the approximative formula
\begin{equation}\label{GG0}
\Gamma_h^{\rm Is}(T)\approx\Gamma_0^{\rm Is}(T),
\end{equation}
where
\begin{equation}\label{G0}
\Gamma_0^{\rm Is}(T)\equiv\frac{h}{T}\cdot\frac{\frac{d\chi(T)}{dT}}{\frac{d^2f_{\rm Is}(T,0)}{dT^2}},
\end{equation}
and, according to \eqref{fIs0} and \eqref{chi},
\begin{equation}\label{dchiT}
\chi(T)=\frac{\beta{\rm e}^{\beta E_{\rm kink}^{\rm Is}}}{4},\qquad
\frac{d\chi(T)}{dT}=-\frac{k_B\beta^2(1+\beta E_{\rm kink}^{\rm Is})}{4}{\rm e}^{\beta E_{\rm kink}^{\rm Is}}.
\end{equation}
At the same time,
\begin{equation}\label{dTT}
T\frac{d^2f_{\rm Is}(T,0)}{dT^2}=-\frac{k_B\beta^2(E_{\rm kink}^{\rm Is})^2{\rm e}^{-\beta E_{\rm kink}^{\rm Is}}}
{(1+{\rm e}^{-\beta E_{\rm kink}^{\rm Is}})^2}.
\end{equation}
The substitutions of \eqref{dchiT} and \eqref{dTT} into \eqref{G0} yield
\begin{equation}\label{G0=}
\Gamma_0^{\rm Is}(T)=h\cdot\frac{(1+\beta E_{\rm kink}^{\rm Is}){\rm e}^{2\beta E_{\rm kink}^{\rm Is}}(1+{\rm e}^{-\beta E_{\rm kink}^{\rm Is}})^2}{4(E_{\rm kink}^{\rm Is})^2}.
\end{equation}
Utilizing in \eqref{dTT} the expression \eqref{fIskink} instead of \eqref{fIs0}, one readily gets
\begin{equation}\label{dTTkink}
T\frac{d^2f_{\rm Is}^{\rm kink}(T,0)}{dT^2}=-h\cdot k_B\beta^2(E_{\rm kink}^{\rm Is})^2{\rm e}^{-\beta E_{\rm kink}^{\rm Is}}
(1-2{\rm e}^{-\beta E_{\rm kink}^{\rm Is}}).
\end{equation}
The replacement of \eqref{dTT} by \eqref{dTTkink} in \eqref{G=} yields
\begin{equation}\label{Gkink}
\Gamma_{\rm kink}^{\rm Is}(T)=h\cdot\frac{(1+\beta E_{\rm kink}^{\rm Is}){\rm e}^{2\beta E_{\rm kink}^{\rm Is}}}
{4(E_{\rm kink}^{\rm Is})^2(1-2{\rm e}^{-\beta E_{\rm kink}^{\rm Is}})}.
\end{equation}

In a similar manner, substituting \eqref{fIsmagn} instead of \eqref{fIs0} into \eqref{G=}, one readily gets
\begin{equation}
\Gamma_{h;\rm magn}^{\rm Is}(T)=\frac{h}{|h|}\cdot
\frac{E_{\rm magn}^{\rm Is}{\rm e}^{-\beta E_{\rm magn}^{\rm Is}}(1-3{\rm e}^{-\beta E_{\rm magn}^{\rm Is}})+2E_{\rm bound}^{\rm Is}{\rm e}^{-\beta E_{\rm bound}^{\rm Is}}}
{(E_{\rm magn}^{\rm Is})^2{\rm e}^{-\beta E_{\rm magn}^{\rm Is}}(1-3{\rm e}^{-\beta E_{\rm magn}^{\rm Is}})+(E_{\rm bound}^{\rm Is})^2{\rm e}^{-\beta E_{\rm bound}^{\rm Is}}}.
\end{equation}

The plots of $\Gamma_h^{\rm Is}(T)$, $\Gamma_{h;\rm magn}^{\rm Is}(T,h)$, $\Gamma_0^{\rm Is}(T)$ and $\Gamma_{h;\rm kink}^{\rm Is}(T)$,
subject to $J_z=3$ and $k_BT=1.3$, are presented in Fig. 2.
The plots of $\Gamma_h^{\rm Is}(T)$, $\Gamma_{h;\rm magn}^{\rm Is}(T,h)$, $\Gamma_0^{\rm Is}(T)$ and $\Gamma_{h;\rm kink}^{\rm Is}(T)$,
subject to $J_z=15$ and $h=0.3$, are presented in Fig. 3.

\section{Summary and discussion}

In the present paper we suggested, that at the vicinity of the magnetization alternation critical point of a {\it gapped} easy-axis XXZ ferromagnet the low temperature thermodynamics is governed by magnons in the both ordered phases \eqref{gg} and by kinks in the QSR \eqref{ll}. The mathematical confirmation of this assumption was given in detail for the Ising model.

It was demonstrated, that the low-temperature asymptotics of the Ising model free energy density \eqref{fIs0} in the ordered phases agrees with the cluster (sector) expansion based on utilization of even one- and two-magnon states \eqref{fIsmagn}. In the QSR the effectiveness of the (kink-based)
cluster (sector) expansion \eqref{fIskink}, \eqref{fIskink12} is less satisfactory, because here the gap is smaller than in the ordered phases, while the temperature is bigger.
At the end of Sect. 2, following the results of \cite{13} and \cite{9}, the strong mathematical evidence is presented
that the same physical picture is inherent in the general easy-axis XXZ chain.

The special attention was given to the $\rm Gr\ddot uneisen$ ratio, for which the two asymptotical expressions has been obtained. The first one, associated
with the ordered phases and corresponding to the regime \eqref{gg}, was derived by the magnon-based cluster (sector) expansion. The second one, associated
with QSR and related to \eqref{ll}, was obtained under the approximate low-$h$ formulas \eqref{GG0} and \eqref{G0}. The corresponding calculations were performed by the combination of the kink-based cluster (sector) expansion and the exact approach. Namely, for the zero-field magnetic
susceptibility in the nominator of the right side of \eqref{G0} we used the exact (well-known) representation, because it is not clear how to get it within the
cluster (sector) expansion. Indeed, this susceptibility is singular at $T=0$ \cite{18}, however the cluster (sector) approach usually results in
regular formulas. The denominator in the right side of \eqref{G0} may be obtained exactly (from \eqref{fIsh=0}) and also approximated
by the kink-based cluster (sector) expansion. It was shown, that under the condition \eqref{lowT}, both the methods give similar results.

We believe that the suggested cluster (sector) expansion also may be applied to the general easy-axis XXZ ferromagnet \cite{13}. This will be useful
for physical interpretation of the results, obtained by the Quantum Transfer Matrix approach \cite{18,19}.

The author is very grateful to S. B. Rutkevich for the helpful discussions.

\appendix
\renewcommand{\theequation}{\thesection.\arabic{equation}}

\section{Direct evaluation of the low-temperature asymptotic for the free energy density}
\setcounter{equation}{0}

Representing \eqref{fIs0} in the form
\begin{equation}\label{fIs}
f_{\rm Is}(T,h)=-\frac{|h|}{2}-\frac{1}{\beta}\log{\Big(\frac{1+{\rm e}^{-\beta|h|}}{2}+
\frac{1-{\rm e}^{-\beta|h|}}{2}
\sqrt{1+\frac{4{\rm e}^{-\beta(J_z+|h|)}}{(1-{\rm e}^{-\beta|h|})^2}}\Big)},
\end{equation}
and accounting for \eqref{gg}, one readily gets
\begin{equation}
\sqrt{1+\frac{4{\rm e}^{-\beta(J_z+|h|)}}{(1-{\rm e}^{-\beta|h|})^2}}
\approx1+\frac{2{\rm e}^{-\beta(J_z+|h|)}}{(1-{\rm e}^{-\beta|h|})^2}
-\frac{{\rm e}^{-2\beta(J_z+|h|)}}{2(1-{\rm e}^{-\beta|h|})^4},
\end{equation}
and
\begin{equation}
\frac{{\rm e}^{-\beta(J_z+|h|)}}{1-{\rm e}^{-\beta|h|}}\approx{\rm e}^{-\beta E_{\rm magn}^{\rm Is}}+{\rm e}^{-\beta E_{\rm bound}^{\rm Is}},\qquad
\frac{{\rm e}^{-2\beta(J_z+|h|)}}{(1-{\rm e}^{-\beta|h|})^3}\approx{\rm e}^{-2\beta E_{\rm magn}^{\rm Is}}.
\end{equation}
Hence,
\begin{eqnarray}\label{A4}
f_{\rm Is}(T,h)\approx-\frac{|h|}{2}-k_BT\log{\Big(1+{\rm e}^{-\beta E_{\rm magn}^{\rm Is}}+{\rm e}^{-\beta E_{\rm bound}^{\rm Is}}
-\frac{{\rm e}^{-2\beta E_{\rm magn}^{\rm Is}}}{4}\Big)}.
\end{eqnarray}
The expansion of the logarithm in \eqref{A4}, according to \eqref{log(1+x)}, yields \eqref{fIsmagn}.

\renewcommand{\theequation}{\thesection.\arabic{equation}}
\section{Plots}
\setcounter{equation}{0}

\begin{figure}[p]
\centering
\includegraphics[width=1.0 \linewidth]{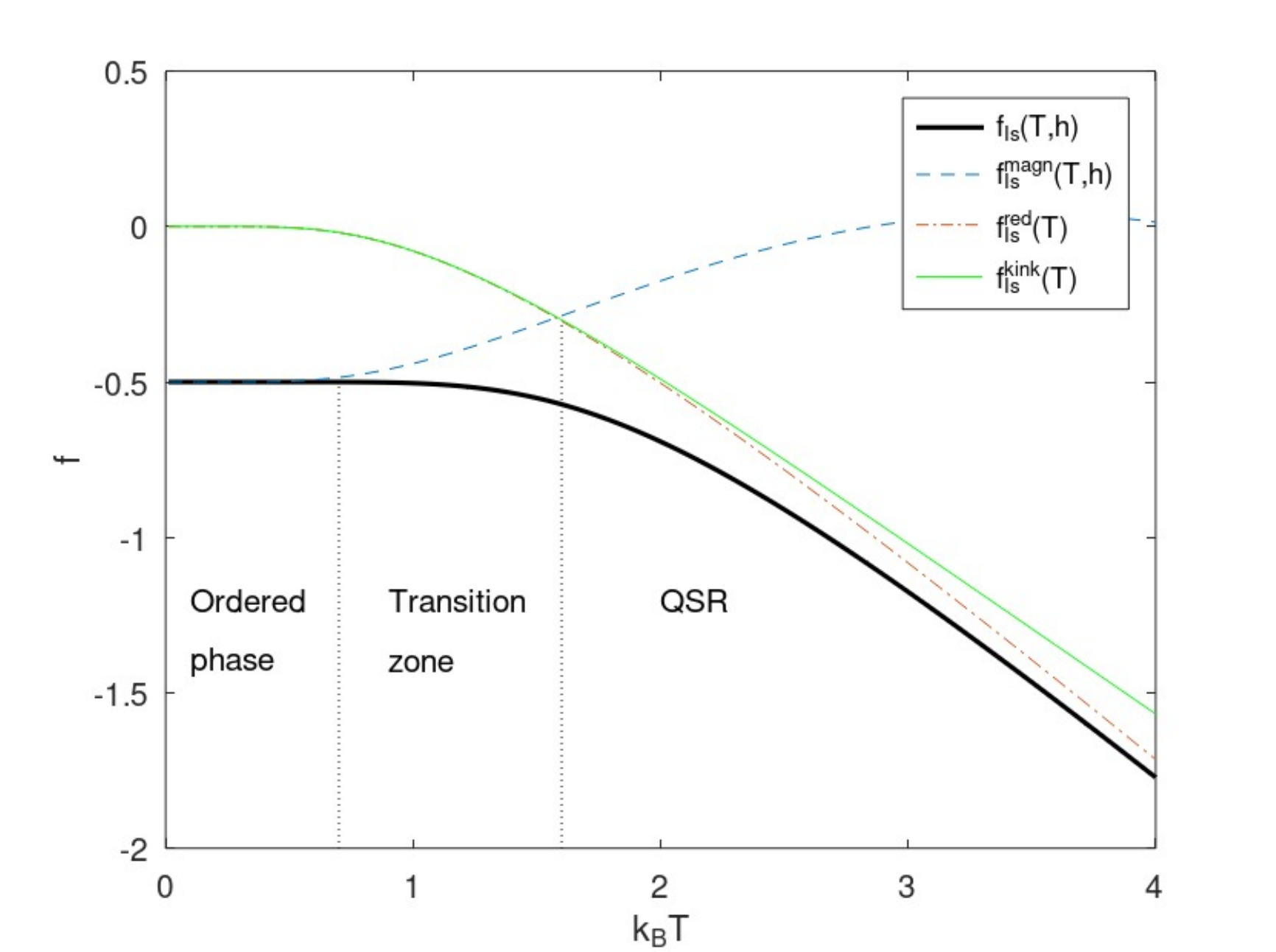}
\caption{$f_{\rm Is}(T,h)$ at $J_z=5$,  $h=1$}
\end{figure}

\begin{figure}[p]
\centering
\includegraphics[width=1.0 \linewidth]{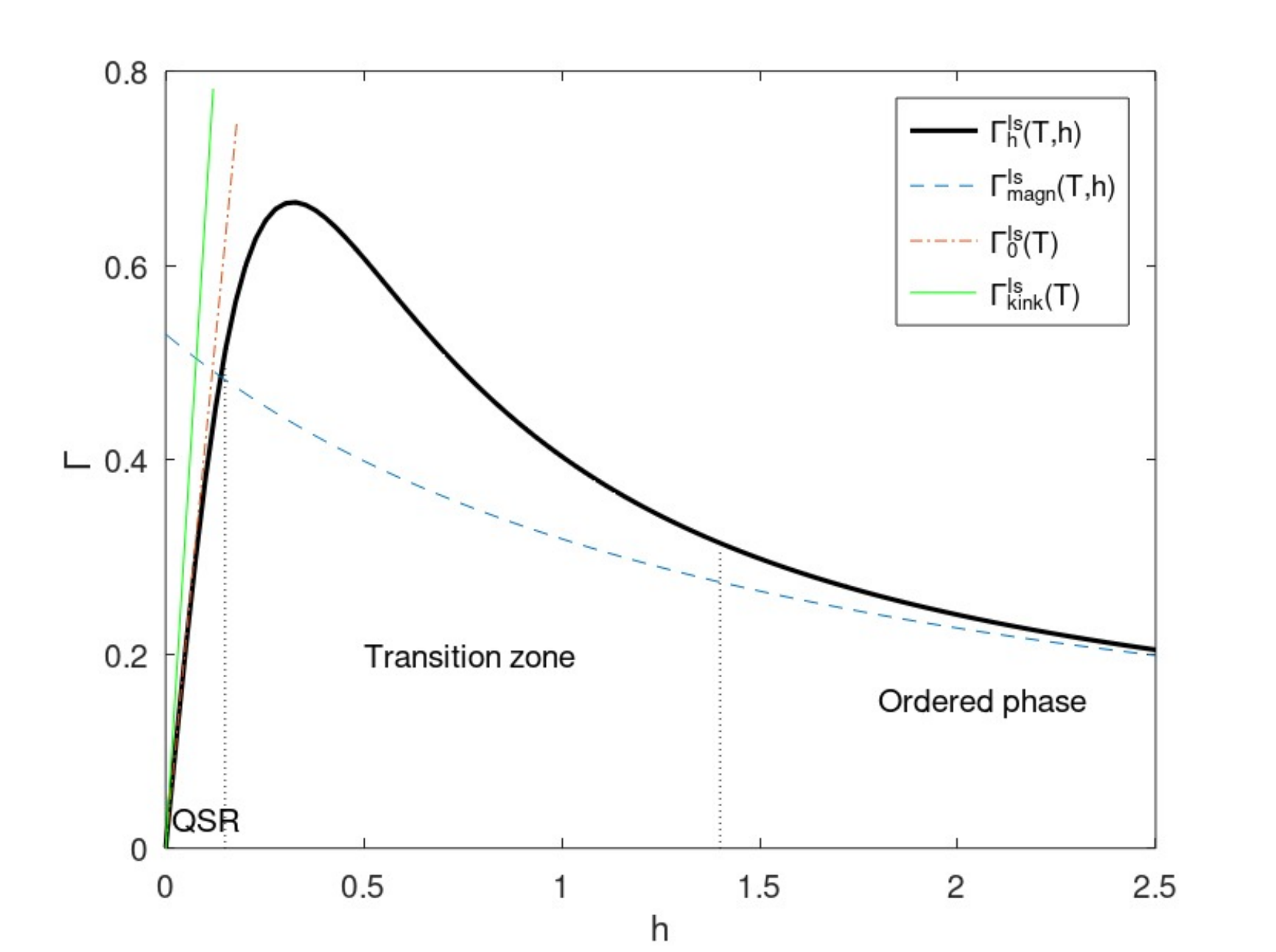}
\caption{$\Gamma_h^{\rm Is}(T)$ at $J_z=3$,  $k_BT=1.3$}
\end{figure}

\begin{figure}[p]
\centering
\includegraphics[width=1.0 \linewidth]{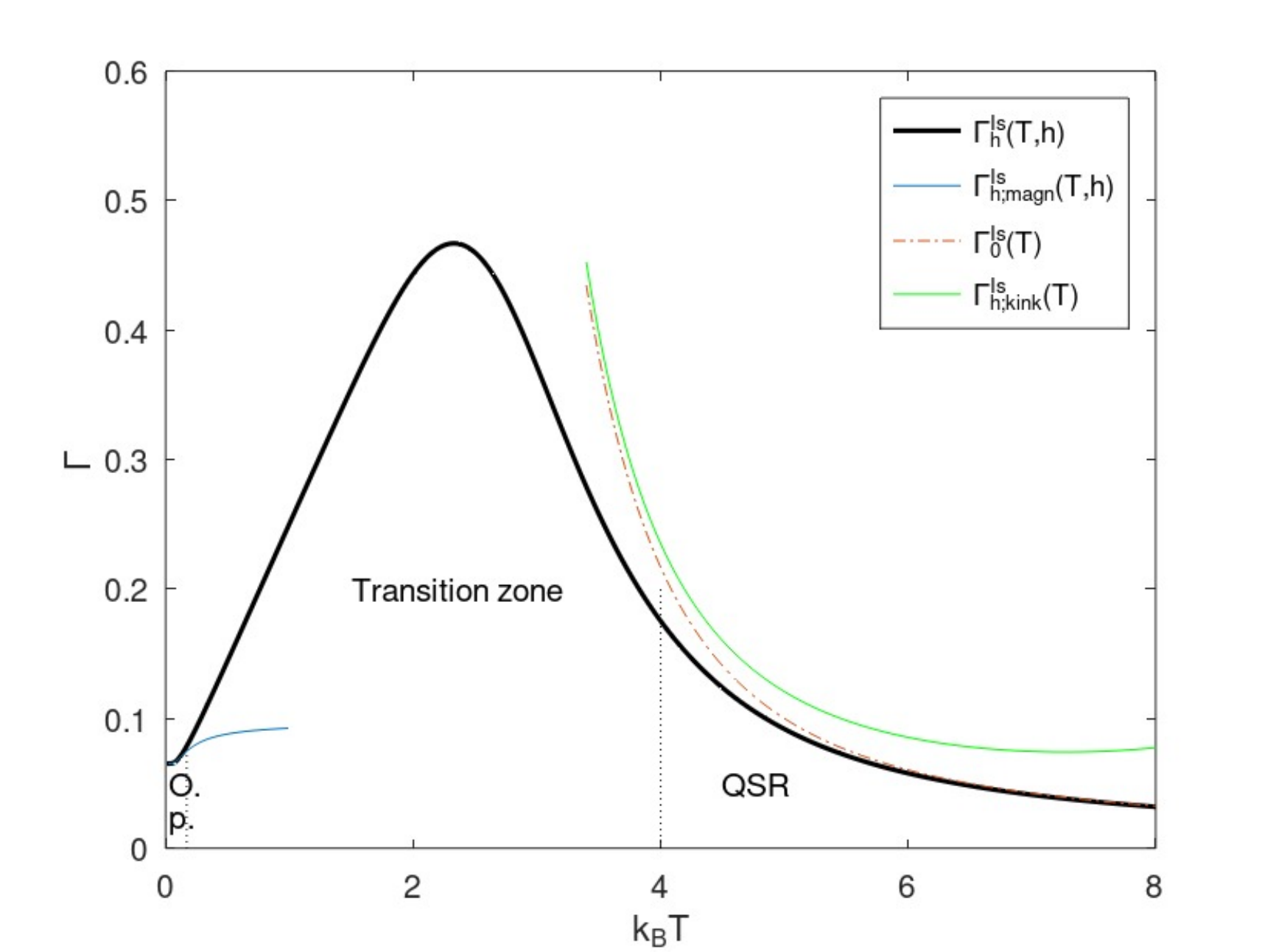}
\caption{$\Gamma_h^{\rm Is}(T)$ at $J_z=15$,  $h=0.3$}
\end{figure}

\end{document}